**Thermodynamics of amide + amine mixtures. 3. Relative permittivities of *N,N*-dimethylformamide + *N*-propylpropan-1-amine, + *N*-butylbutan-1-amine, + butan-1-amine, or + hexan-1-amine systems at several temperatures**

Fernando Hevia, Juan Antonio González*, Isaías García de la Fuente, Luis Felipe Sanz, José Carlos Cobos

G.E.T.E.F., Departamento de Física Aplicada, Facultad de Ciencias, Universidad de Valladolid, Paseo de Belén, 7, 47011 Valladolid, Spain.

*e-mail: jagl@termo.uva.es; Fax: +34-983-423136; Tel: +34-983-423757

# Abstract

Relative permittivities at 1 MHz, $\varepsilon_r$, and at (293.15-303.15) K, are reported for the binary systems *N,N*-dimethylformamide (DMF) + *N*-propylpropan-1-amine (DPA), + *N*-butylbutan-1-amine (DBA), + butan-1-amine (BA) or + hexan-1-amine (HxA). The values of the excess relative permittivities, $\varepsilon_r^E$, have also been determined for these solutions. The measurements were realized by means of a precision impedance analyser 4294A, to which a 16452A cell connected using a 16048G test lead, all of them from Agilent. The $\varepsilon_r^E$ values are large and negative, and diminish when the size of the amine increases along a homologous series, which has been ascribed mainly to the rupture of interactions between like molecules along mixing. Calculations on excess molar orientational polarizabilities support this conclusion, indicating a dominant contribution to $\varepsilon_r^E$ from the orientational polarizability of the molecules in the mixture. The analysis of excess relative Kirkwood correlation factors shows that the correlation between dipoles is effectively decreased along the mixing process.

Keywords: DMF; amine; permittivity, orientational polarizability, Kirkwood correlation factor.

# 1. Introduction

*N,N*-dimethylformamide (DMF) is a very polar compound (3.7 D [1]) able to dissolve many organic substances, as it is an aprotic protophilic substance with excellent donor-acceptor properties. Consequently, DMF has many applications: in the industry [2-4], Nanotechnology [5, 6], or in electrochemical investigation [7, 8], where it is a valuable solvent due to its non-ionic character and high dielectric constant (37.398 at 298.15 K, this work). The knowledge of liquid mixtures containing the amide functional group is also necessary for a deeper understanding of more complex molecules, as those of biological interest [9]. For example, as the chemical environment of proteins is very complex, it is reasonable to pay attention to small molecules similar to the functional groups that constitute the biomolecule. This approach is useful to evaluate interactions in condensed phase environments from thermodynamic, transport, or dielectric properties that are readily available for many organic compounds. On the other hand, it is well known that peptide bond plays a crucial role in the stabilization of secondary structural elements of proteins. The study of the peptide bond is linked to that on the ability of amides to form hydrogen bonds. In fact, amides, in pure state, show a significant local order [10], which, in the case of *N,N*-dialkylamides, has been attributed to the existence of strong dipolar interactions [11], as such compounds are not self-associated.

In contrast, linear primary and secondary amines can form hydrogen bonds, appearing self-associated complexes and even heterocomplexes in mixtures with other associated compounds [12-14]. The amine group is also present in substances of great biological significance. The proteins usually bound to DNA polymers contain various amine groups [15]. Histamine and dopamine are amines with the role of neurotransmitters [14, 16], and the breaking of amino acids releases amines. In addition, the ions of many ionic liquids used in technical applications are related to amines [17].

In a previous study [18], we have reported data of density, $\rho$, speed of sound, $c$, and refractive index, $n_D$, for the binary systems DMF + *N*-propylpropan-1-amine (DPA) or + butan-1-amine (BA) at (293.15-303.15) K, and + *N*-butylbutan-1-amine (DBA) or + hexan-1-amine at 298.15 K. Now, we continue the characterization of these systems providing low-frequency relative permittivities at (293.15-303.15) K. A literature survey shows that $\varepsilon_r$ measurements for DMF + linear primary or secondary amine mixtures are not available. In fact, the $\varepsilon_r$ database for DMF solutions is rather limited. Nevertheless, experimental $\varepsilon_r$ values have been reported for this type of systems including solvents such as: alkanols [19-22], dimethylsulfoxide [23, 24] 1,4-dioxane [23] or propylene glycol [25]. The present $\varepsilon_r$ data, together with our previous $\rho$ and $n_D$ measurements [18] for DMF + amine solutions are used here to determine the corresponding orientational and induced polarizabilities according to the Kirkwood-Fröhlich model [26-28] and the Balankina relative excess Kirkwood correlation factors [29], quantities which are useful to gain insight into the dipole correlations present in the mixtures under consideration.

## 2. Experimental

*Materials*

All the compounds were used without further purification. Information regarding their source and purity is collected in Table 1. Their $\varepsilon_r$ values at 1 MHz and 0.1 MPa are listed in Table 2; they are in good agreement with the literature data.

*Apparatus and procedure*

Binary mixtures have been prepared by mass in small vessels of about 10 cm$^3$, using an analytical balance HR-202 (weighing accuracy 0.01 mg), with all weighings corrected for buoyancy effects. The standard uncertainty in the final mole fraction is estimated to be 0.0001. For the calculation of molar quantities, the relative atomic mass table of 2015 issued by the Commission on Isotopic Abundances and Atomic Weights (IUPAC) [30] was used.

Temperatures were measured using Pt-100 resistances, calibrated according to the ITS-90 scale of temperature, against the triple point of the water and the fusion point of Ga. The repeatability of the equilibrium temperature measurements is 0.01 K, and the corresponding standard uncertainty is 0.02 K.

The $\varepsilon_r$ measurements were realised using a 16452A cell (parallel-plate capacitor) connected, by means of a 16048G test lead, to a precision impedance analyser 4294A; the three of them are from Agilent. The 16452A cell is made of Nickel-plated cobalt (54% Fe, 17% Co, 29% Ni) with a ceramic insulator (alumina, Al$_2$O$_3$). The volume of the sample filling the cell is $\approx 4.8$ cm$^3$. The temperature was controlled by a thermostatic bath LAUDA RE304, with a temperature stability of 0.02 K. Details about the device configuration and calibration are given elsewhere [31]. The relative standard uncertainty of the $\varepsilon_r$ measurements is 0.0001. The relative total uncertainty was estimated to be 0.003 from the differences between our data and values available in the literature for the following pure liquids in the temperature range (288.15–333.15) K: water, benzene, cyclohexane, hexane, nonane, decane, dimethyl carbonate, diethyl carbonate, methanol, 1-propanol, 1-pentanol, 1-hexanol, 1-heptanol, 1-octanol, 1-nonanol and 1-decanol.

## 3. Results

The relative permittivity of an ideal mixture at the same temperature and pressure as the solution under study, $\varepsilon_r^{id}$, is calculated from the expression [32]:

$$\varepsilon_r^{id} = \phi_1 \varepsilon_{r1}^* + \phi_2 \varepsilon_{r2}^* \tag{1}$$

where the volume fraction of component $i$ is defined as $\phi_i = x_i V_{mi}^* / V_m^{id}$. Here, $x_i$ represents the mole fraction of component $i$, $V_{mi}^*$ and $\varepsilon_{ri}^*$ stand for the molar volume and relative permittivity of pure component $i$ respectively, and $V_m^{id} = x_1 V_{m,1}^* + x_2 V_{m,2}^*$ is the ideal molar volume of the mixture at the same temperature and pressure. The excess relative permittivity, $\varepsilon_r^E$, is obtained as

$$\varepsilon_r^E = \varepsilon_r - \varepsilon_r^{id} \tag{2}$$

where $\varepsilon_r$ is the permittivity of the mixture. The necessary volumetric properties were obtained in an earlier work [18].

Volume fractions of DMF, $\phi_1$, $\varepsilon_r$, and $\varepsilon_r^E$ values are listed in Table 3 for DMF (1) + amine (2) systems as functions of the mole fraction of DMF, $x_1$, in the temperature range (293.15 – 303.15) K.

The $\varepsilon_r^E$ data have been fitted by an unweighted linear least-squares regression to a Redlich-Kister equation [33]:

$$\varepsilon_r^E = x_1 (1 - x_1) \sum_{i=0}^{k-1} A_i (2x_1 - 1)^i \tag{3}$$

For each system and temperature, the number, $k$, of necessary coefficients for this regression has been determined by applying an F-test of additional term [34] at 99.5% confidence level. Table 4 includes the parameters $A_i$ obtained, and the standard deviations $\sigma(\varepsilon_r^E)$, defined by:

$$\sigma(\varepsilon_r^E) = \left[ \frac{1}{N-k} \sum_{j=1}^{N} \left( \varepsilon_{r,cal,j}^E - \varepsilon_{r,exp,j}^E \right)^2 \right]^{1/2} \tag{4}$$

where the index $j$ takes one value for each of the $N$ experimental data $\varepsilon_{r,exp,j}^E$, and $\varepsilon_{r,cal,j}^E$ is the corresponding value of the excess property $\varepsilon_r^E$ calculated from eq 3.

## 4. Discussion

Along the present section, the values of the thermophysical properties and the excess functions are referred to $T = 298.15$ K and $\phi_1 = 0.5$.

The value of $\varepsilon_r$ is determined by a number of factors, such as the permanent dipole moment of the molecules, their polarizability or the nature of the liquid structure and collective dynamics. In addition to the fact that DMF has a high molecular dipole moment (3.7 D [1]), its marked polar character is well demonstrated by the relatively high upper critical solution temperatures of DMF + heptane (342.55 K) [35] and DMF + hexadecane (385.15 K) [36] systems. The orientational contribution to $\varepsilon_r$ of strongly polar substances tends to be high and predominant; accordingly, the $\varepsilon_r$ value encountered for DMF is rather large (37.398).

Linear primary and secondary amines are weakly self-associated, which is reflected in the positive $H_m^E(x_1 = 0.5)/\text{J·mol}^{-1}$ values of heptane mixtures: 424 (DPA) [37], 317 (DBA) [37], 1192 (BA) [38], and 962 (HxA) [38]. These values can be ascribed to the disruption of amine-amine interactions along mixing. However, their polarity is rather weak, as shown by their relative permittivities: 3.106 (DPA), 2.903 (DBA), 4.639 (BA), and 3.904 (HxA). It is seen that $\varepsilon_r$ is lower when the amine group is more sterically hindered, i.e., when dipolar interactions are weaker, which suggests that the differences between the permittivity of these amines are mainly of the orientational type (see also the discussion below about molar polarizability).

It is known that the disruption of interactions between like molecules, here the breaking of DMF-DMF and amine-amine interactions, contributes negatively to $\varepsilon_r^E$. The relative $\varepsilon_r$ values of amines show that such contribution decreases when replacing BA by HxA, DPA by DBA, or HxA by DPA. In addition, it is expected that interactions between DMF molecules be more easily broken by those amines with larger aliphatic surfaces. In contrast, the creation of interactions between unlike molecules might have either a positive or a negative contribution to $\varepsilon_r^E$. A positive contribution is encountered when interactions between unlike molecules lead to an increased number of effective dipole moments in the system. Negative contributions arise when interactions between unlike molecules lead to a loss of structure of the liquid with a decrease of the number of effective dipole moments. The mixtures under study are characterized by negative $\varepsilon_r^E$ values (Figures 1 and 2): – 1.372 (DPA), – 1.733 (DBA), – 0.864 (BA), and – 1.262 (HxA). Therefore, the dominant contributions arise from the breaking of interactions between like molecules Interestingly, the $\varepsilon_r^E$ value of the DMF + heptane mixture at $\phi_1 = 0.0171$ and 293.15 K is slightly lower (– 0.24, calculated from data of the literature [39]) than the values of the corresponding systems with amines at the same conditions: – 0.129 (DPA), − 0.146 (DBA), − 0.104 (BA), and − 0.137 (HxA). This reveals that the formation of DMF-amine interactions contributes positively to $\varepsilon_r^E$. On the basis of these considerations, the fact that $\varepsilon_r^E$ (DPA) < $\varepsilon_r^E$ (HxA) can be explained as follows. Firstly, it is possible to assume that the negative contribution to $\varepsilon_r^E$ from the disruption of DMF-DMF interactions is similar in both systems as HxA and DPA have similar aliphatic surfaces. If one takes into account that the negative contribution to $\varepsilon_r^E$ (in absolute value) is larger in the case of HxA, this means that the positive contribution to $\varepsilon_r^E$ due to the creation of interactions between unlike molecules is higher in the case of the HxA mixture. The observed $\varepsilon_r^E$ decrease when replacing BA by HxA, or DPA by DBA, can be explained in similar terms.

In a previous article [18], we reported ultrasonic, volumetric and refractive index (at the wavelength of the sodium D line), $n_D$, data for the mixtures under study. We also evaluated the molar refractivities, which are related to the mean electronic polarizability and dispersive interactions [28, 40]. Using values reported there and $\varepsilon_r$ data determined in this work, the molar orientational polarizabilities (also termed molar orientational polarizations or molar polarizability volumes), $\Pi_m^{or}$, can be calculated. According to the Kirkwood-Fröhlich equation, this quantity can be written as [26-28]

$$\Pi_m^{or} = \frac{N_A \alpha_{or}}{3\varepsilon_0} = \frac{(\varepsilon_r - \varepsilon_{r\infty})(2\varepsilon_r + \varepsilon_{r\infty})}{9\varepsilon_r} V_m \qquad (5)$$

where $\alpha_{or}$ stands for the orientational polarizability (in the case of mixtures, a one-fluid approach is implicit [29]), $\varepsilon_0$ represents the vacuum permittivity, $N_A$ is Avogadro's constant, and $\varepsilon_{r\infty}$ symbolizes the relative permittivity at a frequency at which only the induced (atomic and electronic) polarizability contributes. It has been estimated from the relation [41] $\varepsilon_{r\infty} = 1.1 n_D^2$. Also, the Kirkwood-Fröhlich model provides for the molar induced polarizabilities, $\Pi_m^{ind}$, the following expression:

$$\Pi_m^{ind} = \frac{N_A \alpha_{ind}}{3\varepsilon_0} = \frac{(\varepsilon_{r\infty} - 1)(2\varepsilon_r + \varepsilon_{r\infty})}{9\varepsilon_r} V_m \qquad (6)$$

with $\alpha_{ind}$ meaning the induced polarizability. For the pure compounds, $(\Pi_m^{or})^*/\text{cm}^3 \cdot \text{mol}^{-1}$ = 623.0 (DMF), 39.1 (DPA), 36.8 (DBA), 68.1 (BA), and 64.5 (HxA). On the other hand, $(\Pi_m^{ind})^*/\text{cm}^3 \cdot \text{mol}^{-1}$ = 22.0 (DMF), 47.9 (DPA), 63.0 (DBA), 31.4 (BA), and 45.7 (HxA). The Kirkwood-Fröhlich model then supports our previous statements about the permittivity of the pure compounds, namely: i) $\varepsilon_r$ of DMF arises predominantly from the orientational contribution; and ii) the differences between $\varepsilon_r$ of the pure amines are primarily due to orientational effects, as $(\Pi_m^{or})^*$ and $\varepsilon_r$ vary in the same sense and $(\Pi_m^{ind})^*$ in the opposite.

Using smoothed values for $V_m^E$, $n_D^E$ and $\varepsilon_r^E$ at $\Delta x_1 = 0.01$, we have also determined the excess molar orientational polarizabilities of the mixtures, $(\Pi_m^{or})^E = \Pi_m^{or} - (\Pi_m^{or})^{id}$, as functions of composition, where $(\Pi_m^{or})^{id}$ is calculated substituting the ideal values in eq 5. The results are shown graphically in Figure 3. The values of $(\Pi_m^{or})^E$ / cm$^3$·mol$^{-1}$ are: $-$ 31.6 (DPA), $-$ 41.4 (DBA), $-$ 18.0 (BA) and $-$ 27.8 (HxA). These $(\Pi_m^{or})^E$ values change in line with those of $\varepsilon_r^E$, pointing out that the main contribution to $\varepsilon_r^E$ arises from effects on the orientational polarizability of the molecules. Note the similar shape of the $\varepsilon_r^E$ and $(\Pi_m^{or})^E$ curves (Figures 1-3). The variation of $\Pi_m^{or}$ contrasts with that of the corresponding molar refractivities, whose values increase with the size of the amine [18], indicating that dispersive interactions become more relevant in the same order.

The Balankina relative excess Kirkwood correlation factors [29], $g_{k,\text{rel}}^E = (g_k - g_k^{id})/g_k^{id}$, where $g_k$ and $g_k^{id}$ account respectively for the real and ideal Kirkwood correlation factors, are a useful tool to probe into the structure of the mixtures:

$$g_{k,\text{rel}}^E = \frac{V_m(\varepsilon_r - \varepsilon_{r\infty})(2\varepsilon_r + \varepsilon_{r\infty})\varepsilon_r^{id}(\varepsilon_{r\infty}^{id} + 2)^2}{V_m^{id}(\varepsilon_r^{id} - \varepsilon_{r\infty}^{id})(2\varepsilon_r^{id} + \varepsilon_{r\infty}^{id})\varepsilon_r(\varepsilon_{r\infty} + 2)^2} - 1 \qquad (7)$$

In the ideal mixture, neither correlations between like dipoles are destroyed nor are new correlations between unlike dipoles created. Therefore, the negative $g_{k,\text{rel}}^E$ curves (Figure 4) for DMF + amine systems indicate that the mean correlation between dipoles is destroyed upon mixing. The $g_{k,\text{rel}}^E$ values vary in the same order as the $\varepsilon_r^E$ and $(\Pi_m^{or})^E$ magnitudes: $-$ 0.08 (DPA), $-$ 0.09 (DBA), $-$0.05 (BA), and $-$ 0.07 (HxA), which can be interpreted in similar terms. The minima of the $g_{k,\text{rel}}^E$ curves is reached at lower volume fractions of DMF than in the $\varepsilon_r^E$ and $(\Pi_m^{or})^E$ curves (Table 5). Thus, according to the Kirkwood-Fröhlich model, the destruction of dipole correlations is not the only responsible for the $\varepsilon_r^E$ minima, suggesting the importance of other related effects, such as the number and intensity of interactions created and disrupted upon mixing.

The structure induced in the liquid by the electric field is gradually destroyed as thermal agitation increases. Accordingly, the polarization diminishes with increasing $T$ and the observed values of $(\partial \varepsilon_r^* / \partial T)_p$ are negative. For the pure compounds, $(\partial \varepsilon_r^* / \partial T)_p$/K$^{-1}$ = $-$ 0.175 (DMF), $-$ 0.011 (DPA), $-$ 0.008 (DBA), $-$ 0.019 (BA) and $-$0.013 (HxA). These results point out that the loss of structure is higher when the polar character of the liquid becomes stronger. Similarly, for the DMF mixtures, $(\partial \varepsilon_r / \partial T)_p$/K$^{-1}$ = – 0.090 (DPA), – 0.085 (DBA), – 0.097 (BA) and – 0.091 (HxA). The values of $(\partial \varepsilon_r^E / \partial T)_p$ are slightly positive. The largest value is encountered for the DBA system ($\approx 0.01$ K$^{-1}$). As the DBA is the pure compound which shows the weakest temperature dependence of $\varepsilon_r$, this result suggests that, in the DMF + DBA system, the contribution to $\varepsilon_r$ from the amide-amine interactions with respect to those from the breaking of interactions between like molecules has a more relevant weight when $T$ increases.

## 5. Conclusions

Values of $\varepsilon_r$ and $\varepsilon_r^E$ over the temperature range (293.15-303.15) K have been reported for the systems DMF + DPA, + DBA, + BA or +HxA. The $\varepsilon_r^E$ values are rather large and negative, and decrease when the size of the amine increases along a homologous series. This behaviour has been attributed to a dominant contribution related to the rupture of interactions between like molecules. The mixtures have also been investigated in terms of the excess molar orientational polarizabilities and excess relative Kirkwood correlation factors. This study supports the previous conclusion, pointing out to a dominant contribution to $\varepsilon_r^E$ from the effects on the orientational polarizability of the molecules and an effective loss of correlation between dipoles along mixing.

## Funding

F. Hevia gratefully acknowledges the grant received from the program 'Ayudas para la Formación de Profesorado Universitario (convocatoria 2014), Ministerio de Educación, Cultura y Deporte, Gobierno de España'.

Table 1

Sample description.

| Chemical name | CAS Number | Source | Purification method | Purity[a] | Analysis Method |
|---|---|---|---|---|---|
| *N,N*-dimethylformamide (DMF) | 68-12-2 | Sigma-Aldrich | none | ≥ 0.999 | GC[b] |
| *N*-propylpropan-1-amine (DPA) | 142-84-7 | Fluka | none | ≥ 0.99 | GC[b] |
| *N*-butylbutan-1-amine (DBA) | 111-92-2 | Aldrich | none | ≥ 0.995 | GC[b] |
| butan-1-amine (BA) | 109-73-9 | Sigma-Aldrich | none | ≥ 0.995 | GC[b] |
| hexan-1-amine (HxA) | 111-26-2 | Aldrich | none | ≥ 0.995 | GC[b] |

[a] In mole fraction. [b] Gas chromatography.

Table 2

Relative permittivity, $\varepsilon_r^*$, of pure compounds at temperature $T$, pressure $p$ = 0.1 MPa and frequency $\nu = 1$ MHz. [a]

| Compound | $T$/K | $\varepsilon_r^*$ Exp. | $\varepsilon_r^*$ Lit. |
|---|---|---|---|
| DMF | 293.15 | 38.268 | 38.30 [22] |
|  | 298.15 | 37.398 | 37.65 [42] |
|  | 303.15 | 36.521 | 36.55 [43] |
| DPA | 293.15 | 3.160 | 3.31 [44]<br>3.068 [45] |
|  | 298.15 | 3.106 | 3.24 [44] |
|  | 303.15 | 3.053 | 3.18 [44] |
| DBA | 293.15 | 2.943 | 2.978 [45]<br>2.765 [46] |
|  | 298.15 | 2.903 |  |
|  | 303.15 | 2.863 | 2.697 [46] |
| BA | 293.15 | 4.733 | 4.71 [47]<br>4.88 [45]<br>4.91 [48] |
|  | 298.15 | 4.639 | 4.62 [48] |
|  | 303.15 | 4.546 | 4.57 [47]<br>4.48 [48] |
| HxA | 293.15 | 3.966 | 3.94 [49] |
|  | 298.15 | 3.904 |  |
|  | 303.15 | 3.841 | 3.83 [49] |

[a]The standard uncertainties are: $u(T) = 0.02$ K; $u(p) = 1$ kPa; $u(\nu) = 20$ Hz. The relative standard uncertainty is: $u_r(\varepsilon_r^*) = 0.0001$. The $\varepsilon_r^*$ relative total uncertainty is 0.003.

Table 3

Volume fractions of DMF, $\phi_1$, relative permittivities, $\varepsilon_r$, and excess relative permittivities, $\varepsilon_r^E$, of DMF (1) + amine (2) mixtures as functions of the mole fraction of DMF, $x_1$, at temperature $T$, pressure $p = 0.1$ MPa and frequency $\nu = 1$ MHz. [a]

| $x_1$ | $\phi_1$ | $\varepsilon_r$ | $\varepsilon_r^E$ | $x_1$ | $\phi_1$ | $\varepsilon_r$ | $\varepsilon_r^E$ |
|---|---|---|---|---|---|---|---|
| \multicolumn{8}{c}{DMF (1) + DPA (2) ; $T$/K = 293.15} |
| 0.0000 | 0.0000 | 3.160 |  | 0.5443 | 0.4016 | 15.752 | − 1.507 |
| 0.0604 | 0.0349 | 4.144 | − 0.241 | 0.6001 | 0.4575 | 17.766 | − 1.456 |
| 0.1068 | 0.0630 | 4.914 | − 0.458 | 0.6582 | 0.5197 | 20.054 | − 1.352 |
| 0.1538 | 0.0927 | 5.754 | − 0.661 | 0.6963 | 0.5630 | 21.699 | − 1.227 |
| 0.1874 | 0.1147 | 6.390 | − 0.797 | 0.7563 | 0.6356 | 24.447 | − 1.028 |
| 0.2539 | 0.1605 | 7.773 | − 1.022 | 0.8105 | 0.7062 | 27.175 | − 0.778 |
| 0.3117 | 0.2029 | 9.056 | − 1.227 | 0.8534 | 0.7659 | 29.466 | − 0.583 |
| 0.3921 | 0.2660 | 11.092 | − 1.407 | 0.8993 | 0.8338 | 32.048 | − 0.385 |
| 0.4580 | 0.3220 | 12.962 | − 1.503 | 0.9520 | 0.9177 | 35.221 | − 0.158 |
| 0.5045 | 0.3639 | 14.398 | − 1.538 | 1.0000 | 1.0000 | 38.268 |  |
| \multicolumn{8}{c}{DMF (1) + DPA (2) ; $T$/K = 298.15} |
| 0.0000 | 0.0000 | 3.106 |  | 0.5443 | 0.4014 | 15.380 | − 1.491 |
| 0.0604 | 0.0348 | 4.061 | − 0.238 | 0.6001 | 0.4572 | 17.344 | − 1.440 |
| 0.1068 | 0.0629 | 4.811 | − 0.452 | 0.6582 | 0.5194 | 19.579 | − 1.338 |
| 0.1538 | 0.0926 | 5.633 | − 0.648 | 0.6963 | 0.5627 | 21.181 | − 1.221 |
| 0.1874 | 0.1146 | 6.253 | − 0.783 | 0.7563 | 0.6353 | 23.849 | − 1.043 |
| 0.2539 | 0.1604 | 7.601 | − 1.005 | 0.8105 | 0.7059 | 26.516 | − 0.797 |
| 0.3117 | 0.2027 | 8.851 | − 1.206 | 0.8534 | 0.7657 | 28.762 | − 0.601 |
| 0.3921 | 0.2658 | 10.836 | − 1.385 | 0.8993 | 0.8337 | 31.286 | − 0.409 |
| 0.4580 | 0.3217 | 12.651 | − 1.487 | 0.9520 | 0.9176 | 34.389 | − 0.183 |
| 0.5045 | 0.3637 | 14.069 | − 1.509 | 1.0000 | 1.0000 | 37.398 |  |
| \multicolumn{8}{c}{DMF (1) + DPA (2) ; $T$/K = 303.15} |
| 0.0000 | 0.0000 | 3.053 |  | 0.5443 | 0.4011 | 15.004 | − 1.473 |
| 0.0604 | 0.0348 | 3.974 | − 0.244 | 0.6001 | 0.4569 | 16.914 | − 1.431 |
| 0.1068 | 0.0628 | 4.709 | − 0.446 | 0.6582 | 0.5191 | 19.090 | − 1.336 |
| 0.1538 | 0.0925 | 5.510 | − 0.639 | 0.6963 | 0.5624 | 20.660 | − 1.215 |
| 0.1874 | 0.1145 | 6.112 | − 0.773 | 0.7563 | 0.6350 | 23.252 | − 1.053 |
| 0.2539 | 0.1602 | 7.422 | − 0.993 | 0.8105 | 0.7057 | 25.872 | − 0.799 |
| 0.3117 | 0.2025 | 8.650 | − 1.180 | 0.8534 | 0.7655 | 28.051 | − 0.622 |
| 0.3921 | 0.2656 | 10.585 | − 1.357 | 0.8993 | 0.8335 | 30.513 | − 0.436 |
| 0.4580 | 0.3215 | 12.362 | − 1.451 | 0.9520 | 0.9175 | 33.552 | − 0.208 |
| 0.5045 | 0.3634 | 13.730 | − 1.485 | 1.0000 | 1.0000 | 36.521 | |

DMF (1) + DBA (2) ; $T/K = 293.15$

| | | | | | | | |
|---|---|---|---|---|---|---|---|
| 0.0000 | 0.0000 | 2.943 | | 0.5495 | 0.3558 | 13.618 | − 1.909 |
| 0.0622 | 0.0292 | 3.733 | − 0.243 | 0.5944 | 0.3989 | 15.155 | − 1.896 |
| 0.1117 | 0.0539 | 4.394 | − 0.455 | 0.6458 | 0.4522 | 17.062 | − 1.874 |
| 0.1954 | 0.0991 | 5.628 | − 0.820 | 0.7069 | 0.5220 | 19.694 | − 1.711 |
| 0.2481 | 0.1300 | 6.494 | − 1.047 | 0.7550 | 0.5825 | 21.998 | − 1.547 |
| 0.3044 | 0.1654 | 7.531 | − 1.262 | 0.8058 | 0.6526 | 24.672 | − 1.352 |
| 0.3621 | 0.2045 | 8.691 | − 1.485 | 0.8446 | 0.7111 | 26.989 | − 1.104 |
| 0.3984 | 0.2307 | 9.507 | − 1.595 | 0.8946 | 0.7935 | 30.244 | − 0.764 |
| 0.4462 | 0.2673 | 10.675 | − 1.722 | 0.9517 | 0.8992 | 34.374 | − 0.372 |
| 0.5136 | 0.3235 | 12.533 | − 1.852 | 1.0000 | 1.0000 | 38.311 | |

DMF (1) + DBA (2) ; $T/K = 298.15$

| | | | | | | | |
|---|---|---|---|---|---|---|---|
| 0.0000 | 0.0000 | 2.903 | | 0.5495 | 0.3557 | 13.331 | − 1.850 |
| 0.0622 | 0.0291 | 3.672 | − 0.236 | 0.5944 | 0.3988 | 14.817 | − 1.852 |
| 0.1117 | 0.0538 | 4.314 | − 0.446 | 0.6458 | 0.4521 | 16.687 | − 1.822 |
| 0.1954 | 0.0990 | 5.521 | − 0.799 | 0.7069 | 0.5219 | 19.252 | − 1.666 |
| 0.2481 | 0.1299 | 6.366 | − 1.021 | 0.7550 | 0.5824 | 21.481 | − 1.526 |
| 0.3044 | 0.1653 | 7.378 | − 1.231 | 0.8058 | 0.6525 | 24.118 | − 1.309 |
| 0.3621 | 0.2044 | 8.521 | − 1.438 | 0.8446 | 0.7110 | 26.355 | − 1.091 |
| 0.3984 | 0.2306 | 9.310 | − 1.553 | 0.8946 | 0.7935 | 29.534 | − 0.760 |
| 0.4462 | 0.2672 | 10.445 | − 1.681 | 0.9517 | 0.8992 | 33.603 | − 0.339 |
| 0.5136 | 0.3234 | 12.260 | − 1.806 | 1.0000 | 1.0000 | 37.422 | |

DMF (1) + DBA (2) ; $T/K = 303.15$

| | | | | | | | |
|---|---|---|---|---|---|---|---|
| 0.0000 | 0.0000 | 2.863 | | 0.5495 | 0.3556 | 13.031 | − 1.803 |
| 0.0622 | 0.0291 | 3.610 | − 0.233 | 0.5944 | 0.3987 | 14.481 | − 1.804 |
| 0.1117 | 0.0538 | 4.237 | − 0.437 | 0.6458 | 0.4520 | 16.306 | − 1.774 |
| 0.1954 | 0.0990 | 5.412 | − 0.784 | 0.7069 | 0.5218 | 18.800 | − 1.629 |
| 0.2481 | 0.1299 | 6.236 | − 1.000 | 0.7550 | 0.5823 | 20.971 | − 1.495 |
| 0.3044 | 0.1653 | 7.225 | − 1.203 | 0.8058 | 0.6524 | 23.534 | − 1.292 |
| 0.3621 | 0.2043 | 8.338 | − 1.403 | 0.8446 | 0.7109 | 25.713 | − 1.082 |
| 0.3984 | 0.2305 | 9.112 | − 1.511 | 0.8946 | 0.7934 | 28.815 | − 0.758 |
| 0.4462 | 0.2671 | 10.218 | − 1.637 | 0.9517 | 0.8991 | 32.785 | − 0.346 |
| 0.5136 | 0.3233 | 11.986 | − 1.761 | 1.0000 | 1.0000 | 36.528 | |
| DMF (1) + BA (2) ; $T$/K = 293.15 | | | | | | | |
| 0.0000 | 0.0000 | 4.733 | | 0.4992 | 0.4362 | 18.425 | − 0.926 |
| 0.0467 | 0.0366 | 5.756 | − 0.204 | 0.5956 | 0.5334 | 21.782 | − 0.827 |
| 0.1016 | 0.0807 | 6.990 | − 0.447 | 0.6458 | 0.5860 | 23.618 | − 0.754 |
| 0.1505 | 0.1209 | 8.192 | − 0.593 | 0.6990 | 0.6432 | 25.663 | − 0.626 |
| 0.2030 | 0.1651 | 9.522 | − 0.744 | 0.7895 | 0.7443 | 29.231 | − 0.446 |
| 0.2442 | 0.2005 | 10.603 | − 0.849 | 0.8476 | 0.8119 | 31.649 | − 0.293 |
| 0.3014 | 0.2509 | 12.198 | − 0.943 | 0.8959 | 0.8698 | 33.690 | − 0.193 |
| 0.3552 | 0.2995 | 13.811 | − 0.959 | 0.9490 | 0.9353 | 36.001 | − 0.077 |
| 0.4466 | 0.3852 | 16.683 | − 0.959 | 1.0000 | 1.0000 | 38.246 | |
| DMF (1) + BA (2) ; $T$/K = 298.15 | | | | | | | |
| 0.0000 | 0.0000 | 4.639 | | 0.4992 | 0.4359 | 17.983 | − 0.926 |
| 0.0467 | 0.0366 | 5.632 | − 0.205 | 0.5956 | 0.5331 | 21.269 | − 0.822 |
| 0.1016 | 0.0806 | 6.840 | − 0.438 | 0.6458 | 0.5856 | 23.068 | − 0.741 |
| 0.1505 | 0.1207 | 8.004 | − 0.586 | 0.6990 | 0.6429 | 25.059 | − 0.626 |
| 0.2030 | 0.1649 | 9.305 | − 0.732 | 0.7895 | 0.7441 | 28.564 | − 0.434 |
| 0.2442 | 0.2003 | 10.363 | − 0.833 | 0.8476 | 0.8117 | 30.905 | − 0.306 |
| 0.3014 | 0.2506 | 11.917 | − 0.926 | 0.8959 | 0.8696 | 32.912 | − 0.194 |
| 0.3552 | 0.2992 | 13.487 | − 0.947 | 0.9490 | 0.9352 | 35.165 | − 0.089 |
| 0.4466 | 0.3848 | 16.277 | − 0.959 | 1.0000 | 1.0000 | 37.375 | |
| DMF (1) + BA (2) ; $T$/K = 303.15 | | | | | | | |
| 0.0000 | 0.0000 | 4.546 | | 0.4992 | 0.4355 | 17.533 | − 0.916 |
| 0.0467 | 0.0365 | 5.504 | − 0.207 | 0.5956 | 0.5327 | 20.735 | − 0.817 |
| 0.1016 | 0.0805 | 6.687 | − 0.429 | 0.6458 | 0.5853 | 22.487 | − 0.744 |
| 0.1505 | 0.1206 | 7.807 | − 0.589 | 0.6990 | 0.6425 | 24.428 | − 0.629 |
| 0.2030 | 0.1647 | 9.076 | − 0.728 | 0.7895 | 0.7438 | 27.855 | − 0.436 |
| 0.2442 | 0.2000 | 10.120 | − 0.811 | 0.8476 | 0.8115 | 30.162 | − 0.290 |
| 0.3014 | 0.2503 | 11.632 | − 0.905 | 0.8959 | 0.8695 | 32.127 | − 0.177 |
| 0.3552 | 0.2989 | 13.152 | − 0.936 | 0.9490 | 0.9351 | 34.330 | − 0.068 |
| 0.4466 | 0.3845 | 15.865 | − 0.956 | 1.0000 | 1.0000 | 36.470 | |
| DMF (1) + HxA (2) ; $T$/K = 293.15 | | | | | | | |
| 0.0000 | 0.0000 | 3.966 | | 0.5571 | 0.4226 | 17.096 | − 1.369 |
| 0.0559 | 0.0333 | 4.841 | − 0.267 | 0.6043 | 0.4705 | 18.792 | − 1.316 |
| 0.1089 | 0.0664 | 5.734 | − 0.510 | 0.6439 | 0.5127 | 20.311 | − 1.245 |
| 0.1672 | 0.1046 | 6.807 | − 0.748 | 0.7060 | 0.5829 | 22.872 | − 1.093 |
| 0.2106 | 0.1344 | 7.678 | − 0.899 | 0.7438 | 0.6282 | 24.506 | − 1.013 |
| 0.2553 | 0.1663 | 8.627 | − 1.045 | 0.7954 | 0.6935 | 26.946 | − 0.813 |
| 0.3031 | 0.2020 | 9.710 | − 1.186 | 0.8467 | 0.7627 | 29.497 | − 0.636 |
| 0.3520 | 0.2402 | 10.936 | − 1.271 | 0.8948 | 0.8319 | 32.073 | − 0.435 |
| 0.4062 | 0.2847 | 12.386 | − 1.348 | 0.9508 | 0.9183 | 35.291 | − 0.181 |
| 0.4504 | 0.3229 | 13.644 | − 1.400 | 1.0000 | 1.0000 | 38.275 | |
| 0.5074 | 0.3748 | 15.427 | − 1.398 | | | | |

DMF (1) + HxA (2) ; $T$/K = 298.15

| | | | | | | | |
|---|---|---|---|---|---|---|---|
| 0.0000 | 0.0000 | 3.904 | | 0.5571 | 0.4225 | 16.715 | − 1.351 |
| 0.0559 | 0.0333 | 4.759 | − 0.261 | 0.6043 | 0.4704 | 18.362 | − 1.309 |
| 0.1089 | 0.0664 | 5.629 | − 0.501 | 0.6439 | 0.5126 | 19.846 | − 1.240 |
| 0.1672 | 0.1046 | 6.681 | − 0.729 | 0.7060 | 0.5828 | 22.334 | − 1.105 |
| 0.2106 | 0.1343 | 7.523 | − 0.883 | 0.7438 | 0.6281 | 23.949 | − 1.008 |
| 0.2553 | 0.1662 | 8.455 | − 1.020 | 0.7954 | 0.6934 | 26.304 | − 0.842 |
| 0.3031 | 0.2019 | 9.518 | − 1.153 | 0.8467 | 0.7626 | 28.833 | − 0.633 |
| 0.3520 | 0.2401 | 10.701 | − 1.251 | 0.8948 | 0.8319 | 31.328 | − 0.460 |
| 0.4062 | 0.2846 | 12.113 | − 1.331 | 0.9508 | 0.9183 | 34.469 | − 0.215 |
| 0.4504 | 0.3228 | 13.350 | − 1.374 | 1.0000 | 1.0000 | 37.423 | |
| 0.5074 | 0.3747 | 15.080 | − 1.384 | | | | |

DMF (1) + HxA (2) ; $T$/K = 303.15

| | | | | | | | |
|---|---|---|---|---|---|---|---|
| 0.0000 | 0.0000 | 3.841 | | 0.5571 | 0.4223 | 16.326 | − 1.317 |
| 0.0559 | 0.0333 | 4.675 | − 0.254 | 0.6043 | 0.4702 | 17.927 | − 1.281 |
| 0.1089 | 0.0663 | 5.526 | − 0.482 | 0.6439 | 0.5124 | 19.376 | − 1.211 |
| 0.1672 | 0.1045 | 6.549 | − 0.707 | 0.7060 | 0.5826 | 21.803 | − 1.079 |
| 0.2106 | 0.1343 | 7.371 | − 0.859 | 0.7438 | 0.6279 | 23.372 | − 0.990 |
| 0.2553 | 0.1662 | 8.278 | − 0.995 | 0.7954 | 0.6932 | 25.663 | − 0.833 |
| 0.3031 | 0.2018 | 9.315 | − 1.121 | 0.8467 | 0.7625 | 28.132 | − 0.629 |
| 0.3520 | 0.2400 | 10.470 | − 1.215 | 0.8948 | 0.8318 | 30.585 | − 0.441 |
| 0.4062 | 0.2845 | 11.844 | − 1.295 | 0.9508 | 0.9183 | 33.628 | − 0.225 |
| 0.4504 | 0.3227 | 13.047 | − 1.340 | 1.0000 | 1.0000 | 36.523 | |
| 0.5074 | 0.3745 | 14.739 | − 1.341 | | | | |

[a]The standard uncertainties are: $u(T) = 0.02$ K; $u(p) = 1$ kPa; $u(\nu) = 20$ Hz; $u(x_1) = 0.0001$; $u(\phi_1) = 0.0002$. The relative standard uncertainty is $u_r(\varepsilon_r) = 0.0001$. The relative combined standard uncertainty is $U_{rc}(\varepsilon_r^E) = 0.03$.

Table 4

Coefficients $A_i$ and standard deviations, $\sigma(\varepsilon_r^E)$ (eq 4), for the representation of $\varepsilon_r^E$ at temperature $T$ and pressure $p = 0.1$ MPa for DMF (1) + amine (2) systems by eq 3.

| System | $T$/K | $A_0$ | $A_1$ | $A_2$ | $A_3$ | $\sigma(\varepsilon_r^E)$ |
|---|---|---|---|---|---|---|
| DMF + DPA | 293.15 | − 6.12 | − 0.40 | 2.54 | 1.2 | 0.011 |
| | 298.15 | − 6.03 | − 0.45 | 2.25 | 0.9 | 0.010 |
| | 303.15 | − 5.92 | − 0.57 | 1.93 | 0.8 | 0.012 |
| DMF + DBA | 293.15 | − 7.36 | − 3.16 | 1.5 | 1.3 | 0.013 |
| | 298.15 | − 7.17 | − 3.12 | 1.37 | 1.3 | 0.011 |
| | 303.15 | − 6.97 | − 3.01 | 1.14 | 1.1 | 0.011 |
| DMF + BA | 293.15 | − 3.76 | 1.72 | 0.5 | | 0.014 |
| | 298.15 | − 3.73 | 1.66 | 0.48 | | 0.008 |
| | 303.15 | − 3.71 | 1.64 | 0.60 | | 0.010 |
| DMF + HxA | 293.15 | − 5.61 | 0.32 | 1.09 | | 0.011 |
| | 298.15 | − 5.53 | 0.15 | 0.90 | | 0.006 |
| | 303.15 | − 5.39 | 0.09 | 0.82 | | 0.007 |

Table 5

Compositions of DMF ($x_1$, mole fraction; $\phi_1$, volume fraction) and values of the minima of the excess relative permittivity, $\varepsilon_r^E$, excess molar orientational polarizability, $(\Pi_m^{or})^E$, and relative excess Kirkwood correlation factors, $g_{k,\text{rel}}^E$ (eq. 7), curves at temperature $T = 298.15$ K and pressure $p = 0.1$ MPa for DMF (1) + amine (2) systems.

| System | $\varepsilon_r^E$ minimum | | | $(\Pi_m^{or})^E$ minimum | | | $g_{k,\text{rel}}^E$ minimum | | |
|---|---|---|---|---|---|---|---|---|---|
| | $x_1$ | $\phi_1$ | $\varepsilon_r^E$ | $x_1$ | $\phi_1$ | $(\Pi_m^{or})^E/\text{cm}^3\cdot\text{mol}^{-1}$ | $x_1$ | $\phi_1$ | $g_{k,\text{rel}}^E$ |
| DMF + DPA | 0.51 | 0.37 | −1.509 | 0.46 | 0.32 | −37.5 | 0.22 | 0.14 | −0.15 |
| DMF + DBA | 0.58 | 0.38 | −1.858 | 0.51 | 0.32 | −50.2 | 0.26 | 0.14 | −0.18 |
| DMF + BA | 0.41 | 0.35 | −0.972 | 0.40 | 0.34 | −20.5 | 0.20 | 0.16 | −0.09 |
| DMF + HxA | 0.49 | 0.36 | −1.383 | 0.44 | 0.31 | −33.4 | 0.22 | 0.14 | −0.13 |

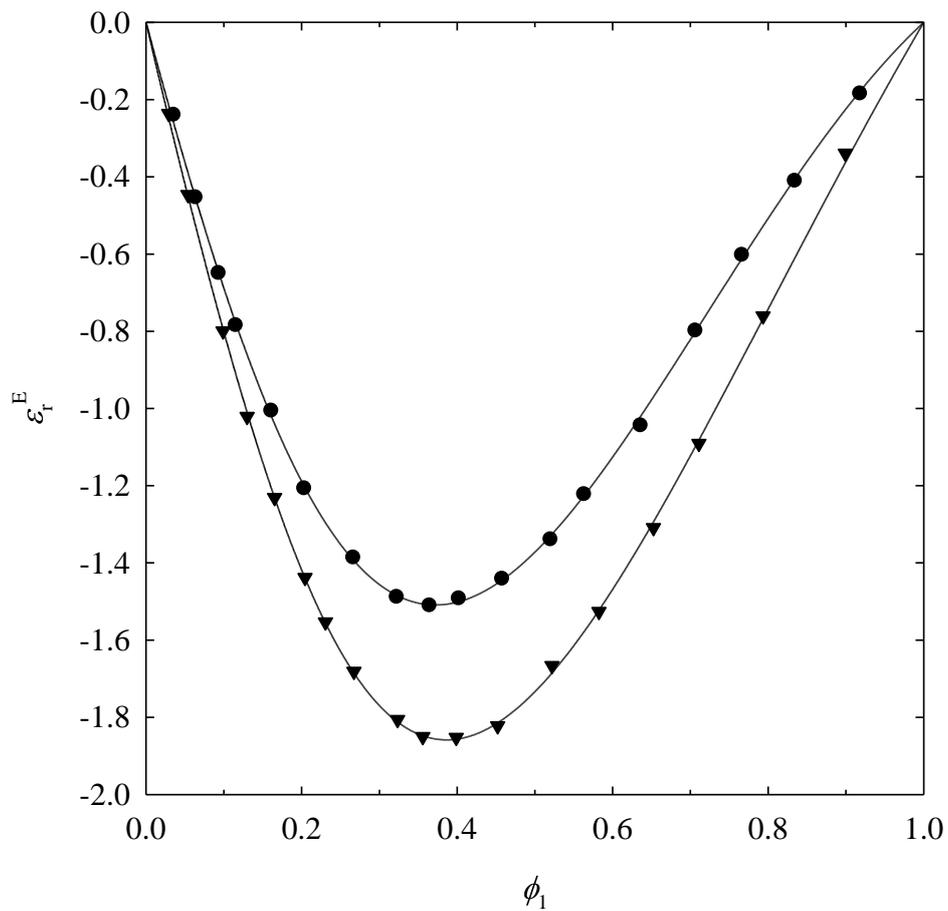

Figure 1

Excess relative permittivities, $\varepsilon_r^E$, for DMF (1) + DPA (2), or + DBA (2) systems at 0.1 MPa, 298.15 K and 1 MHz. Full symbols, experimental values (this work): (●), DPA; (▼), DBA. Solid lines, calculations with eq 3 using the coefficients from Table 4.

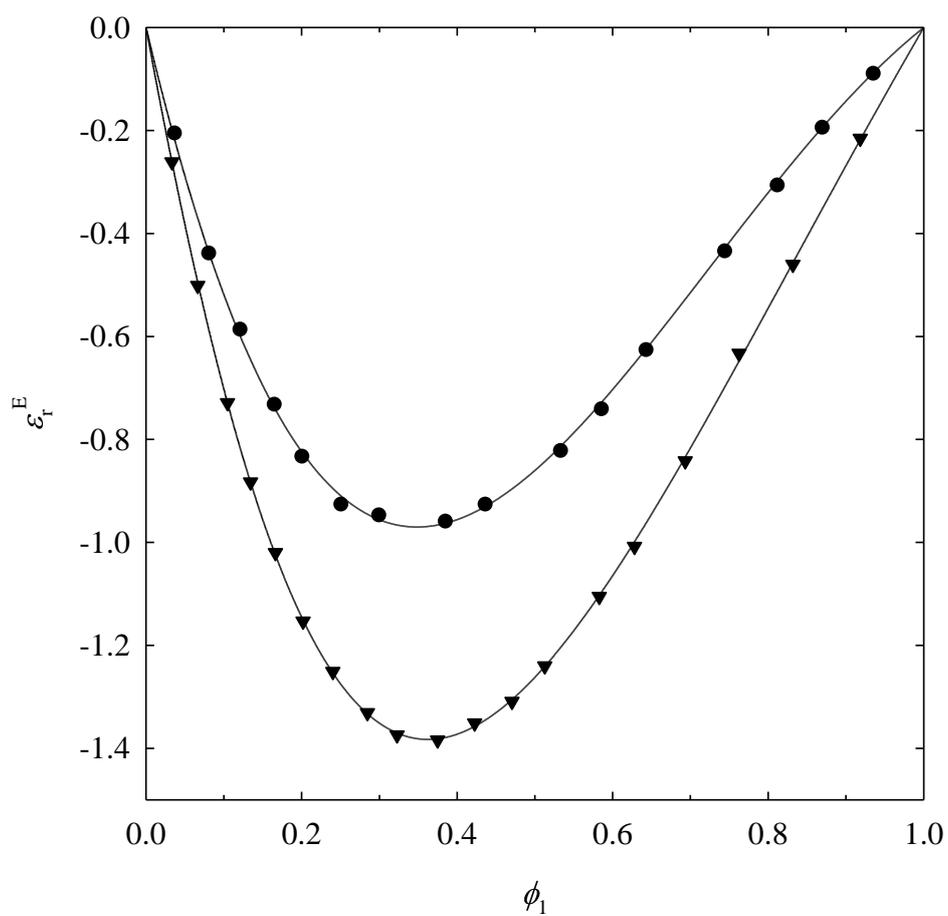

Figure 2

Excess relative permittivities, $\varepsilon_r^E$, for DMF (1) + BA (2), or + HxA (2) systems at 0.1 MPa, 298.15 K and 1 MHz. Full symbols, experimental values (this work): (●), BA; (▼), HxA. Solid lines, calculations with eq 3 using the coefficients from Table 4.

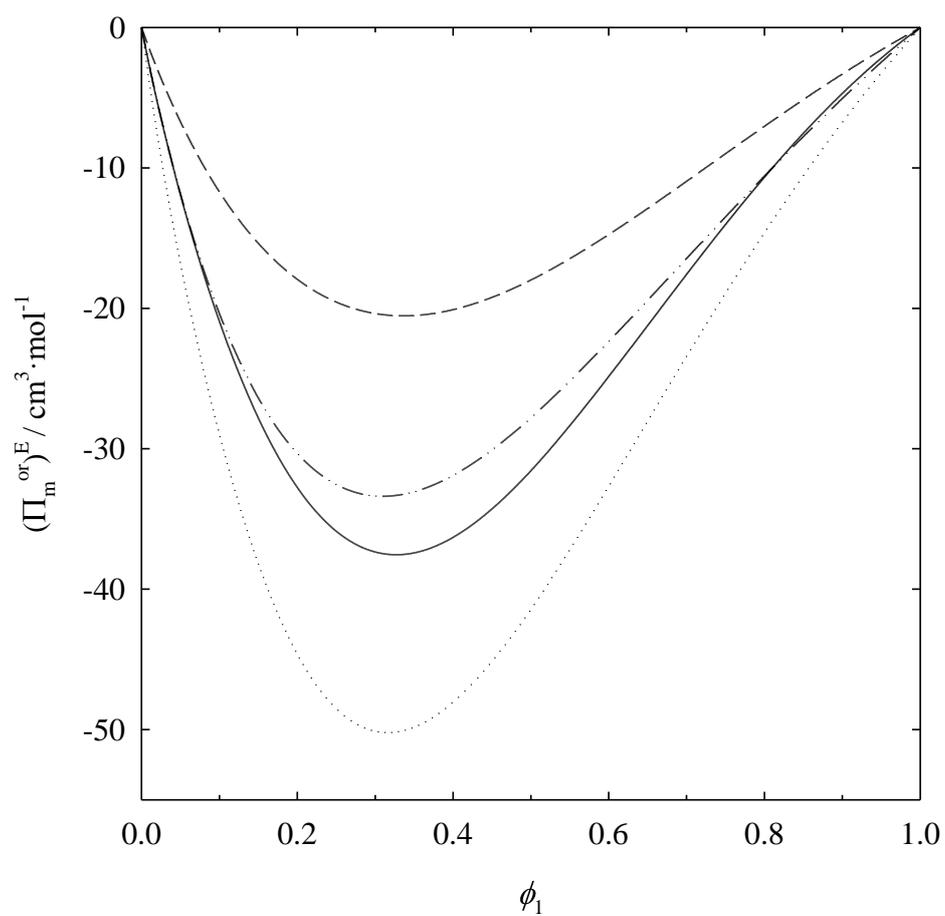

Figure 3

Excess molar orientational polarizability, $(\Pi_m^{or})^E$, for DMF (1) + amine (2) systems at 0.1 MPa and 298.15 K: DPA (——); DBA (······); BA (– – – –); HxA (–··–··–).

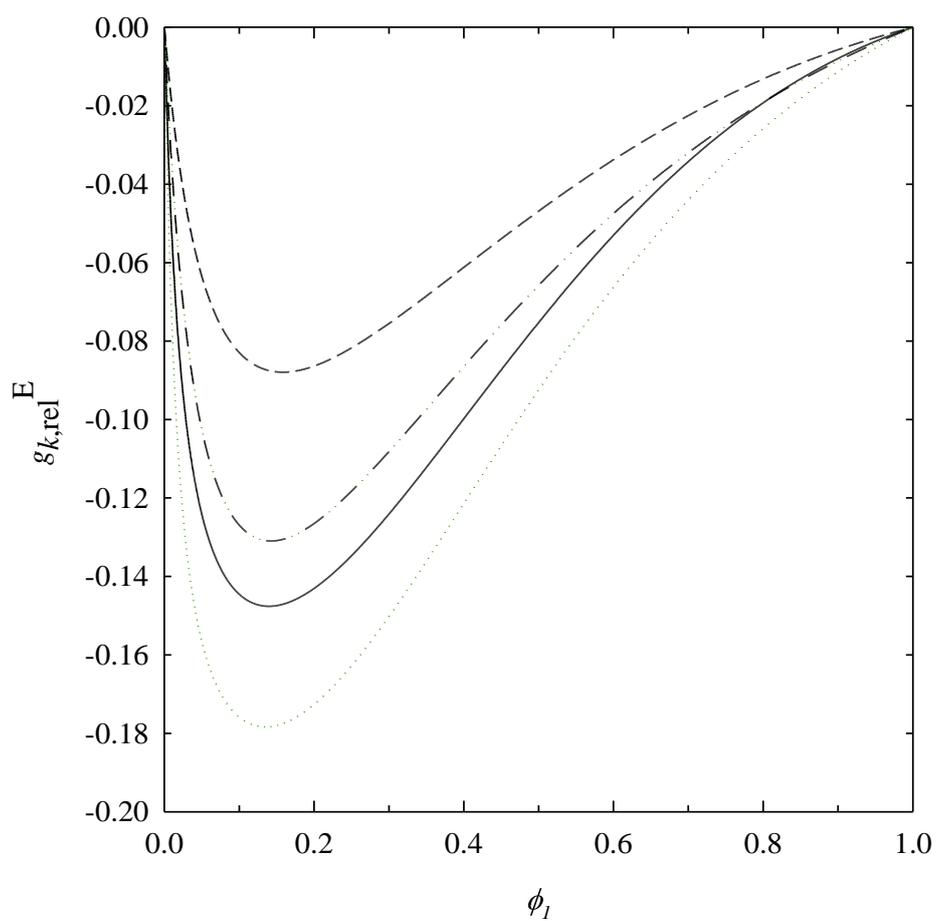

Figure 4

Relative excess Kirkwood correlation factors, $g_{k,\text{rel}}^{\text{E}}$, for DMF (1) + amine (2) systems at 0.1 MPa and 298.15 K: DPA (——); DBA (······); BA (– – – –); HxA (–··–··–).